# Survival of a brown dwarf after engulfment by a red giant star

P.F.L. Maxted[1], R. Napiwotzki[2], P.D. Dobbie[3], M.R. Burleigh[3]

**Many sub-stellar companions (usually planets but also some brown dwarfs) have been identified orbiting solar-type stars. These stars can engulf their sub-stellar companions when they become red giants. This interaction may explain several outstanding problems in astrophysics[1-5] but is poorly understood, e.g., it is unclear under which conditions a low mass companion will evaporate, survive the interaction unchanged or gain mass.[1,4,5] Observational tests of models for this interaction have been hampered by a lack of positively identified remnants, i.e., white dwarf stars with close, sub-stellar companions. The companion to the pre-white dwarf AA Doradus may be a brown dwarf, but the uncertain history of this star and the extreme luminosity difference between the components make it difficult to interpret the observations or to put strong constraints on the models.[6,7] The magnetic white dwarf SDSS J121209.31+013627.7 may have a close brown dwarf companion[8] but little is known about this binary at present. Here we report the discovery of a brown dwarf in a short period orbit around a white dwarf. The properties of both stars in this binary can be directly observed and show that the brown dwarf was engulfed by a red giant but that this had little effect on it.**

[1] Astrophysics Group, Keele University, Keele, Staffordshire, ST5 5BG, UK.

[2] Centre for Astrophysics Research, University of Hertfordshire, College Lane, Hatfield AL10 9AB, UK.

[3] Department of Physics and Astronomy, University of Leicester, University Road, Leicester, LE1 7RH, UK

WD0137-349 (BPS CS 29504-0036) was first noted as an unremarkable faint, blue star in a survey for metal poor stars.[9] Improved spectroscopy showed it to be a white dwarf.[10] We obtained high resolution spectra of WD0137-349 as part of the SPY programme to identify the progenitors of Type Ia supernovae.[11] We noticed features in these spectra due to a low mass companion in a close orbit so we obtained further observations (shown in Figure 1) to measure the orbital period and mass ratio (Table 1). From the mass of the white dwarf (Table 2) and the mass ratio we derive a mass for the companion of $(0.053\pm0.006)M_\odot$ (solar masses). This mass is well below the limit of about $0.075M_\odot$ commonly used to distinguish stars from brown dwarfs.[12] Brown dwarfs, by definition, are not massive enough to support core hydrogen burning but do undergo a brief phase of deuterium burning soon after they form. They then start to cool, so the spectral type of a brown dwarf, which is a measure of its temperature, is also a measure of its age. The observed infrared flux distribution of WD0137-349 (Figure 2) is consistent with a model of an old brown dwarf companion with a mass of $0.055M_\odot$ but inconsistent with models for companions that are young brown dwarfs or stars. The orbital period of WD0137-349 is approximately 116 minutes and the stars are separated by only $0.65R_\odot$ (solar radii).

Brown dwarf companions to white dwarfs are rare – less than 0.5% of white dwarfs have a brown dwarf companion at any separation.[13] However, there are many white dwarfs that are known to have a low mass star as a companion in a short period orbit. In cataclysmic variable stars (CVs) mass transfer onto the white dwarf from the companion through the inner Lagrangian point produces strong emission lines in the optical spectrum. No such lines are seen in our data for WD0137-349 so we conclude that it is not a CV. There are several good candidates for sub-stellar companions in CVs, but the extra light due to accretion makes it difficult to confirm their masses.[14,15]

Short period white dwarfs binaries with low mass companions that do not transfer mass are known as pre-CVs because the loss of orbital angular momentum by gravitational wave radiation (GWR) and other mechanisms will result in the shrinkage of the orbit, the initiation of mass transfer and the formation of a CV.[16] The timescale for WD0137-349 to become a CV through the loss of GWR is about 1.4Ga, at which time the orbit period will be 60 – 80 minutes. This is close to the minimum orbital period seen in CVs. This raises the possibility that some fraction of CVs with very low mass companions may have formed as the result of the evolution of binaries like WD0137-349, rather than by extensive mass loss from the companion. A simulation of the population of pre-CVs formed from binaries like WD0137-349 shows that most of these binaries will have evolved from a solar-type star with a brown dwarf companion separated by a few au.[17] Although a few such systems have been found[18], such binaries are known to be rare[19] so it is likely that the contribution of stars like WD0137-349 to the total CV population is a few percent or less.

The simulation of the population of binaries like WD0137-349 assumes that white dwarfs with close, low mass companions are the result of "common envelope evolution". In this scenario, the more massive star in a binary system becomes a red giant once it has exhausted hydrogen in its core. The red giant will interact with its companion when its radius becomes comparable to the separation of the binary. The details of the interaction are uncertain but some low mass companions will be engulfed by the red giant, i.e., the core of the red giant and the low mass companion share a common envelope. If the companion is not sufficiently massive to force the envelope to co-rotate with its orbit, the drag on the companion will cause it to quickly spiral in towards the core of the red giant. Some fraction of the orbital energy released, $\alpha_{CE}$, will be deposited as kinetic energy in the envelope, which is ejected from the binary system. The radius of a red giant is determined principally by the mass of its core. This is effectively the mass of the resulting white dwarf, so we can calculate the value of $\alpha_{CE}$ required to explain the formation of WD0137-349 assuming a range of red giant masses, $M_g$. The lowest possible value of $M_g$ is 0.8 $M_\odot$ because stars less massive than this do not evolve to the red giant stage within the lifetime of the Galaxy. This provides a lower limit of $\alpha_{CE} \approx 0.6$. This value of $\alpha_{CE}$ is similar to that derived for pre-CVs with more massive companions.[20] If the red giant was more massive than $1.25 M_\odot$ the value of $\alpha_{CE}$ required exceeds 1. These values can be compared directly to the results of simulations of the common envelope phase, although no such simulations for systems resembling WD0137-349 are available to us at present. Simple physical arguments suggest that low mass companions to red giants will be evaporated during the common envelope phase if they are less massive than some limit $m_{crit}$. The value of $m_{crit}$ is uncertain, but is expected to be about $0.02 M_\odot$.[1,4,5] The properties of WD0137-349 show that $m_{crit}$ is at most $0.05 - 0.06 M_\odot$. The mass of WD0137-349 $(0.4 M_\odot)$ is lower than the typical mass of a single white dwarf $(0.6 M_\odot)$, as is expected for systems in which a common envelope phase has prematurely removed the envelope of a red giant.[21]

Some models predict that planets may accrete a substantial fraction of the mass in the red giant envelope prior to a common envelope phase, resulting in the formation of a binary with similar properties to WD0137-349.[4] Since most of the current mass of the brown dwarf was accreted from the red giant in this scenario, its spectral type would imply an age similar to that of the white dwarf, i.e. about 250 Ma in the case of WD0137-349. The spectral type in this case is expected to be $\approx$L1-2 ($T_{eff} \approx$ 2200K). In contrast, if the brown dwarf has always been close to its current mass, by the time the common-envelope phase occurs the brown dwarf will have been cooling for the lifetime of the solar-type star (giga-years). The common envelope phase proceeds on a dynamical timescale of a few years, which is negligible when compared to this thermal timescale so very little mass or heat can be gained by the brown dwarf in this phase. The spectral type of the brown dwarf is then expected to be in the much cooler T-dwarf range ($T_{eff} <$1500K). The companion may appear to be slightly hotter than this because it intercepts about 1% of the light from the white dwarf. The asymmetric heating and rotation of the brown dwarf will produce a small but detectable modulation of the brightness at infrared wavelengths of WD0137-349 on the orbital period ("reflection effect"). An accurate measurement of the intrinsic spectral type and luminosity of the brown dwarf will therefore require infrared spectroscopy and photometry at a range of orbital phases to determine and account for the irradiation from the white dwarf. Despite this complication, the existing infrared photometry (Figure 2) is more consistent with a spectral type for the brown dwarf slightly earlier (hotter) than T5, rather than with a spectral type of L1-2. Therefore, the existing data favour the scenario in which WD0137-349 formed by a common envelope phase which had little effect on the brown dwarf, rather than by accretion onto a planet.

**Acknowledgements** Based on observations collected at the European Southern Observatory, Chile. This publication makes use of data products from the Two Micron All Sky Survey (2MASS). PDD is supported by PPARC. RN and MB acknowledge the support of PPARC Advanced Fellowships.

**Author Contributions** P.M. analysed and interpreted the data from which the orbital period and other properties of the system have been measured. R.N. identified WD0137-349 as a strong candidate for a brown dwarf companion to a white dwarf from observations obtained as part of the SPY programme, of which is the principal investigator. M.B. and P.D. analysed the 2MASS data for this object. All authors discussed the results and commented on the manuscript.

**Author information** Correspondence and requests for material should be addressed to P.M. (pflm@astro.keele.ac.uk).

**Competing interests statement** Authors declare they have no competing financial interests

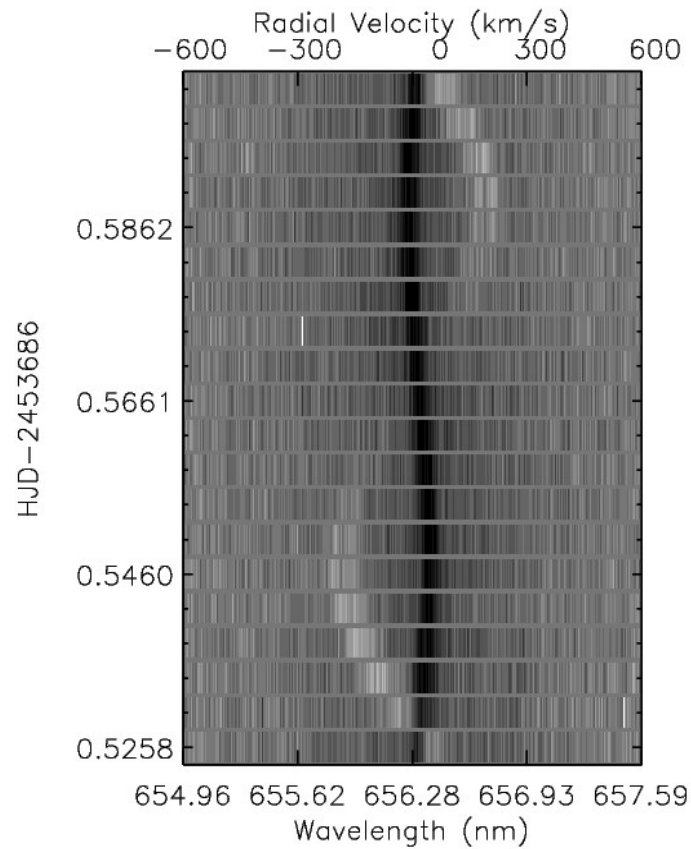

**Figure 1 Trailed spectrograms of WD0137-349.** Spectra were obtained with the Ultraviolet–Visual Echelle Spectrograph (UVES) mounted on ESO's Kueyen telescope. The radial velocity is indicated assuming a rest wavelength of 656.276nm. The time of observation is indicated on the y-axis. The exposure time are indicated by vertical extent of each spectrogram. The spectra have been normalized so the continuum value is 1. The grey-scale representation is a linear scale from 0.4 (black) to 1.4 (white). The sharp absorption feature is the $H_\alpha$ line due to absorption by hydrogen in the atmosphere of the white dwarf star. The sinusoidal change in wavelength is due to the Doppler shift of the white dwarf as it orbits in a binary system. The emission feature seen moving in anti-phase to the absorption line arises in the atmosphere of the low mass

companion to the white dwarf. The variation in the strength and width of this line show that it is produced by irradiation of one hemisphere of the companion by the white dwarf. These observations are available from the ESO Science Archive Facility (http://archive.eso.org, programmes 276.D-5014 and 167.D-0407).

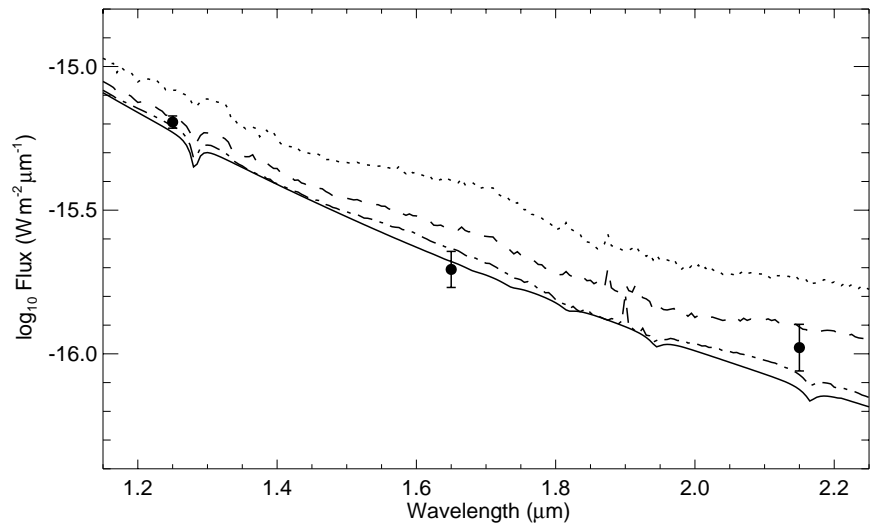

**Figure 2 Infrared flux distribution of WD0137-349.** Measurements from the 2MASS archive (solid circles, with 1σ error bars) are compared to a synthetic white dwarf spectrum from a pure-hydrogen model atmosphere normalized using the observed V band magnitude (black, solid line). Also shown are the synthetic white dwarf spectrum combined with spectra of known brown dwarf stars scaled to the appropriate distance as follows (top-to-bottom): L0 (dotted line), L4 (dashed line), T2 (dashed-dotted line).[22] The orbital phase at which these data were obtained is unknown, so no account has been made for heating of the companion by the white dwarf.

**Table 1: Spectroscopic orbit of WD0137-349**

| | |
|---|---|
| $P$ (days) | 0.0803 ± 0.0002 |
| $T_0$ (Heliocentric Julian Date) | 2453686.5276 ± 0.0001 |
| $K_1$ (km/s) | 27.9 ± 0.3 |
| $K_2$ (km/s) | -187.5 ± 1.1 |
| $\gamma_1$ (km/s) | 17.8 ± 0.3 |
| $\gamma_2$ (km/s) | 3.4 ± 1.0 |
| Correction to $K_2$ (km/s) | 21 ± 7 |
| $m_1 \sin^3 i$ ($M_\odot$) | 0.097 ± 0.008 |
| $m_2 \sin^3 i$ ($M_\odot$) | 0.013 ± 0.001 |
| $a \sin i$ ($R_\odot$) | 0.375 ± 0.014 |
| Mass ratio ($m_2/m_1 = K_1/K_2$) | 0.134 ± 0.006 |

The measured radial velocities of the $H_\alpha$ absorption line at time T are given by $\gamma_1 + K_1 \sin(2\pi[T-T_0]/P)$, and similarly for the emission line ($P$ is orbital period, $T_0$ is reference time, $\gamma_1$ and $\gamma_2$ are the apparent mean radial velocities, $K_1$ and $K_2$ are the semi-amplitudes of the spectroscopic orbits). A correction to the value of $K_2$ has been applied because the light in the emission line we measured is offset from the centre of the companion towards the centre-of-mass of the binary.[23] Accounting for this effect will increase the value of $K_2$ by some fraction of the projected rotational velocity of the brown dwarf, $v_{rot} \sin i$, where $i$ is the inclination of the orbital plane to the plane of the sky. The mass of the white dwarf is $m_1$ and the mass of the brown dwarf is $m_2$. We estimate $i \approx 35°$ from the mass of the white dwarf given in Table 1 and the value of $m_1 \sin^3 i$. Strong tidal forces will ensure that the brown dwarf rotates synchronously so for a typical brown dwarf radius of $0.1R_\odot$ we obtain $v_{rot}\sin i$ = 35 km/s. The minimum masses, $m_1 \sin^3 i$ and $m_2 \sin^3 i$, and the projected separation ($a\sin i$) are calculated using Kepler's Laws from $K_1$, $K_2$ and $P$ including the correction to $K_2$ described above. The difference $\gamma_1 - \gamma_2$ = (14.4 ± 1.1) km/s is due mainly to the gravitational redshift of the white dwarf and agrees well with the value of (13.3 ± 1.1) km/s implied by the parameters in Table 1.

### Table 2: Properties of the white dwarf WD0137-349

| | |
|---|---|
| Apparent visual magnitude, V | 15.33 ± 0.02[9] |
| Effective temperature,$T_{eff}$ (K) | 16,500 ± 500 |
| Surface gravity , *log g* (c.g.s. units) | 7.49 ± 0.08 |
| Mass ($M_\odot$) | 0.39 ± 0.035 |
| Age (Ma) | 250 ± 80 |
| Luminosity ($L_\odot$) | 0.023 ± 0.004 |
| Radius ($R_\odot$) | 0.0186 ± 0.0012 |
| Distance (parsecs) | 102 ± 3 |

The effective temperature ($T_{eff}$) and surface gravity (g) are derived from an analysis of the hydrogen absorption lines in the optical spectrum of WD0137-349 using the technique described in ref. 24. The mass and age are inferred from $T_{eff}$ and log g using models of white dwarfs with a range of compositions.[25,26] The age is the time since formation of the white dwarf by ejection of the red giant envelope. The distance is inferred from the luminosity, apparent visual magnitude and a model atmosphere for hydrogen rich white dwarfs.[27] The main factors that determine the white dwarf mass are well understood and have been tested against observations, i.e., pure hydrogen model atmospheres for moderately hot white dwarfs and the mass-radius relation for degenerate stars.[28]